\documentclass[10pt]{article}

\usepackage{geometry}
\usepackage{amsmath,amssymb,amsfonts}

\usepackage{xcolor}
\definecolor{graycolor}{gray}{0.9}
\newcommand{\revised}[1]{#1} 
\usepackage{microtype}
\usepackage{setspace} 
\usepackage[utf8]{inputenc}
\usepackage[english]{babel}
\usepackage{times}
\usepackage{array}
\usepackage{soul}
\usepackage{cite}
\usepackage[numbers,sort&compress]{natbib}
\usepackage{natbib}

\usepackage{titlesec} 
\titleformat {\section} [block] {\raggedright \fontsize{10}{10}\selectfont\bfseries} {\thesection. \space} {0pt} {}
\titlespacing {\section} {0pt} {12pt} {6pt}
\titleformat {\subsection} [block] {\raggedright \fontsize{10}{10}\selectfont\itshape} {\thesubsection .\space} {0pt} {}
\titlespacing {\subsection} {0pt} {12pt} {6pt}
\titleformat {\subsubsection} [block] {\raggedright \fontsize{10}{10}\selectfont} {\thesubsubsection .\space} {0pt} {}
\titlespacing {\subsubsection} {0pt} {12pt} {6pt}
\titleformat {\paragraph} [block] {\raggedright \fontsize{10}{10}\selectfont} {} {0pt} {}
\titlespacing {\paragraph} {0pt} {12pt} {6pt}

\usepackage{array} \newcommand{\PreserveBackslash}[1]{\let\temp=\\#1\let\\=\temp}
\newcolumntype{C}[1]{>{\PreserveBackslash\centering}m{#1}}
\newcolumntype{R}[1]{>{\PreserveBackslash\raggedleft}m{#1}}
\newcolumntype{L}[1]{>{\PreserveBackslash\raggedright}m{#1}}
\usepackage{lineno}
\usepackage{tabularx}
\usepackage{colortbl}
\usepackage{graphicx}
\usepackage{float}
\usepackage[export]{adjustbox}
\usepackage{caption}
\usepackage{fancyhdr} 
\usepackage{lastpage}
\usepackage{layout}
\usepackage{setspace} 
\usepackage{enumitem}
\usepackage{booktabs}
\usepackage{arydshln}
\usepackage{multirow}
\usepackage{color}
\usepackage{hyperref} 
\hypersetup{
	colorlinks=true,
	linkcolor=blue,
	filecolor=blue,
	urlcolor=black,
	citecolor=cyan,
}

\fancypagestyle{firstpage}{
    \setlength{\headsep}{2.2cm}
    
    \setlength{\footskip}{1.5cm}
    \fancyhf{}
    \lhead{\begin{table}[H]
        \centering
        \begin{tabular}{L{2.5cm}C{10cm}C{3.1cm}R{2cm}}
            \includegraphics[scale=0.035]{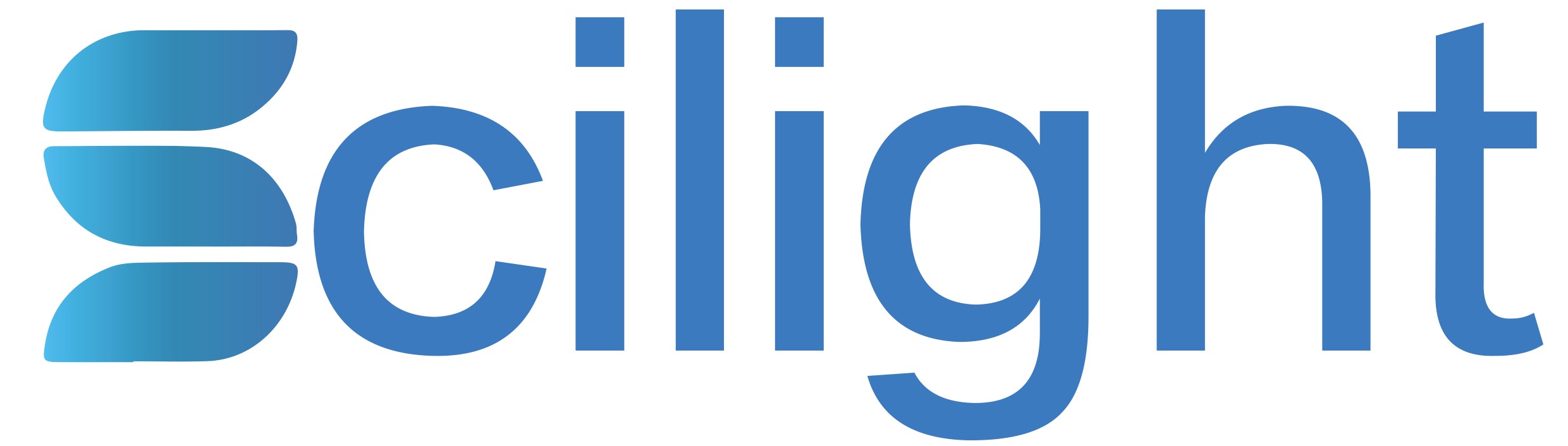} \vspace{-6pt}& \cellcolor{graycolor}\begin{tabular}[c]{@{}c@{}}\textit{Physics and the Cosmos}\\ \href{https://www.sciltp.com/journals/pac}{https://www.sciltp.com/journals/pac}\end{tabular} & \includegraphics[scale=0.022]{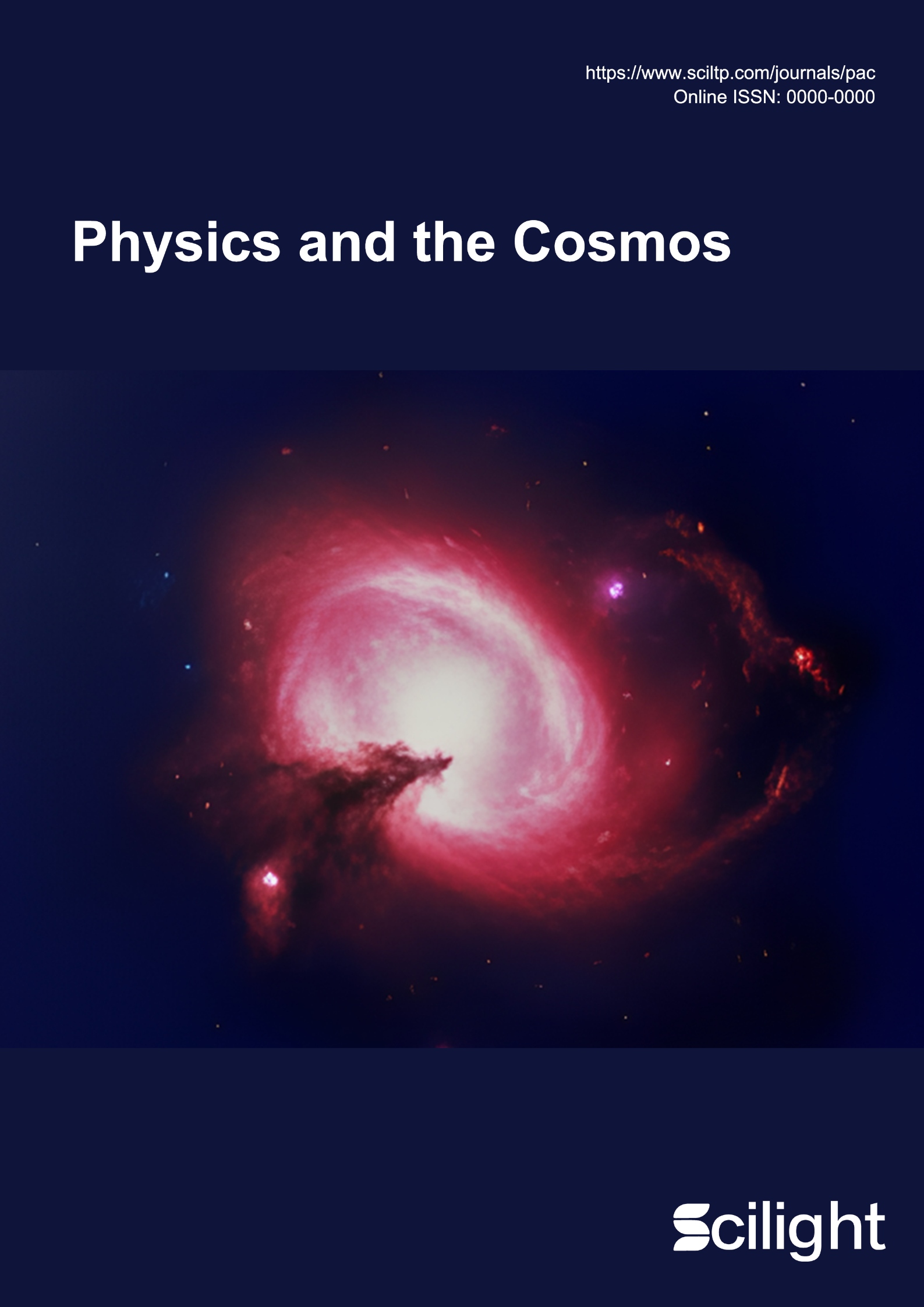} \vspace{-3pt}\\
        \end{tabular}
        \vspace{-22pt}
    \end{table}}
   
    \fancyfoot[C]{
        \vspace{-1.55cm}
        \begin{table}[H]
            \begin{minipage}[c]{0.15\columnwidth}
                \includegraphics[scale=0.5]{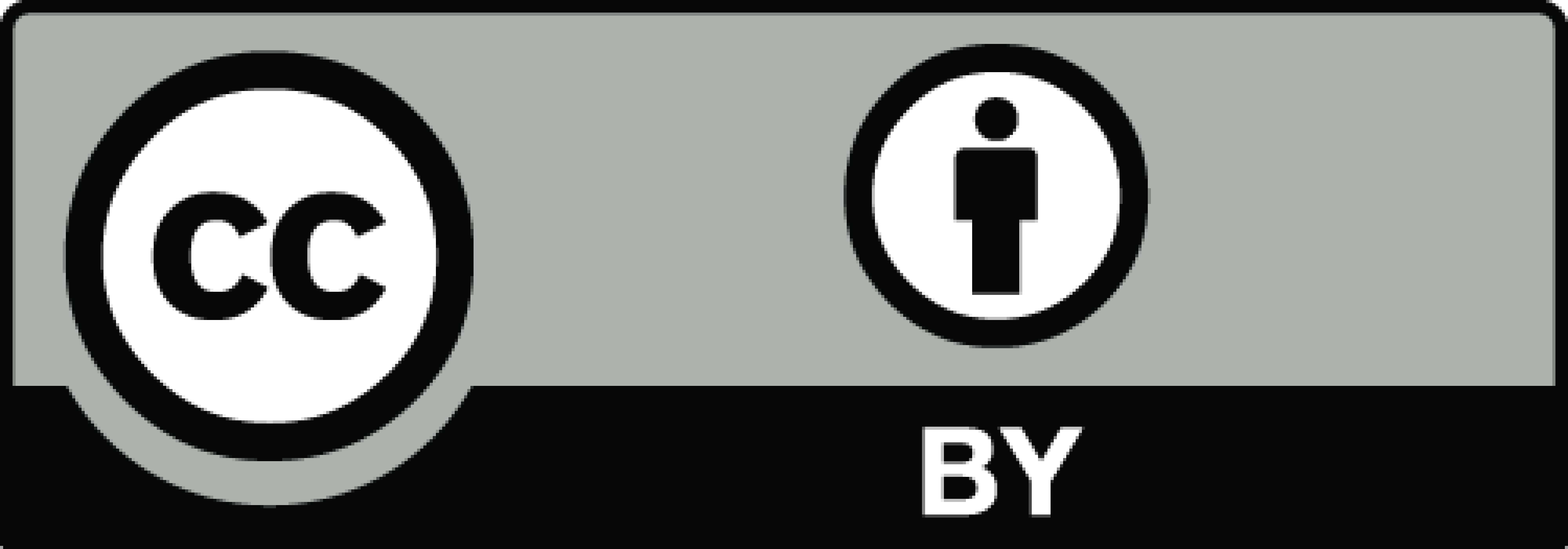} \vspace{1.1pt}
            \end{minipage}
            \hfill
            \begin{minipage}[c]{0.85\columnwidth}
                \scriptsize \textbf{Copyright:} © 2025 by the authors. This is an open access article under the terms and conditions of the Creative Commons Attribution (\mbox{CC BY}) license  (\href{https://creativecommons.org/licenses/by/4.0/}{https://creativecommons.org/licenses/by/4.0/}). \\ \textbf{Publisher’s Note:} Scilight stays neutral with regard to jurisdictional claims in published maps and institutional affiliations.
            \end{minipage}
    \end{table}}
    \vspace{-0.55cm}
}

\usepackage{aas_macros}

\graphicspath{{figures/}}

\begin{document}

\newgeometry{left=2.5cm, right=2.5cm, top=1.8cm, bottom=4cm}
	\thispagestyle{firstpage}
	\nolinenumbers
	{\noindent \textit{Article}}
	\vspace{4pt} \\
	{\fontsize{18pt}{10pt}\textbf{A Population View of the Cosmic-Ray Knee: The Role of Variance in Supernova Maximum Rigidities}}
	\vspace{16pt} \\
	{\large Carmelo Evoli\textsuperscript{1,2}}
	\vspace{6pt}
	 \begin{spacing}{0.9}
		{\noindent \small
			\textsuperscript{1}	\parbox[t]{0.98\linewidth}{Gran Sasso Science Institute (GSSI), Viale Francesco Crispi 7, 67100 L’Aquila, Italy} \\
			\textsuperscript{2}	INFN-Laboratori Nazionali del Gran Sasso (LNGS), via G. Acitelli 22, 67100 Assergi (AQ), Italy \\
		    {*} \parbox[t]{0.98\linewidth}{Correspondence: carmelo.evoli@gssi.it} \\\\
		\footnotesize \textbf{How To Cite}: Evoli, C. A Population View of the Cosmic-Ray Knee: The Role of Variance in Supernova Maximum Rigidities. \emph{Physics and the Cosmos} \textbf{2026}, \emph{1}(1), 8. \href{https://doi.org/10.53941/pac.2026.100008}{https://doi.org/10.53941/pac.2026.100008}}.\\
	\end{spacing}

\begin{table}[H]
\noindent\rule[0.15\baselineskip]{\textwidth}{0.5pt} 
\begin{tabular}{lp{12cm}}  
 \small 
  \begin{tabular}[t]{@{}l@{}} 
  \footnotesize  Received: 11 May 2026 \\
  \footnotesize  Revised: 15 June 2026 \\
   \footnotesize Accepted: 10 August 2026 \\
  \footnotesize  Published: 13 August 2026
  \end{tabular} &
  \textbf{Abstract:}
The broad shape of the Galactic cosmic-ray knee challenges source models in which all supernova remnants share a nearly universal, sharp maximum rigidity. We investigate whether the knee can instead arise as a population effect, produced by source-to-source variations in the maximum energy of Galactic supernova remnants. We derive the population-averaged spectrum for sources with sharp individual cutoffs and distributed $E_{\max}$, showing that it is given by an underlying propagated power law multiplied by the survival probability of the cutoff distribution. A lognormal distribution of $E_{\max}$ naturally produces a smooth, continuously curving knee, while a power-law tail gives an approximately constant post-knee steepening. We then connect the lognormal width to supernova-remnant physics through maximum-energy scalings with explosion energy and ambient density, finding that the expected variance is mainly driven by the spread in explosion energies. Fitting the measured proton spectrum with a two-component lognormal-cutoff model, we find that the PeV component requires $\sigma_{\log_{10}E_{\max}}\simeq 0.24$. \revised{This width is substantially smaller than the variance expected for the full Galactic remnant population, indicating that the PeV component must originate from a more restricted and comparatively homogeneous subset of remnants.} Our results show that the knee can be understood as the gradual exhaustion of a heterogeneous population of PeV-capable supernova remnants, without requiring a universal maximum rigidity.
\\
\\
  & 
  \textbf{Keywords:} Astroparticle Physics; Cosmic Rays; Galactic Supernovae
\end{tabular}
\noindent\rule[0.15\baselineskip]{\textwidth}{0.5pt} 
\end{table}


\section{Introduction}
\label{sec:introduction}

The origin of the Galactic cosmic-ray knee remains one of the central open questions in high-energy astrophysics. 
The steepening of the all-particle spectrum at a few PeV has long been interpreted either as a signature of the maximum rigidity reached by Galactic accelerators, as the result of a change in the efficiency with which cosmic rays escape from the Galaxy, or as some combination of the two~\citep{Hoerandel:2002yg}. 
In both cases, the knee carries direct information on the highest-rigidity end of the Galactic cosmic-ray population.

\begin{figure}[t]
\centering
\includegraphics[width=0.75\textwidth]{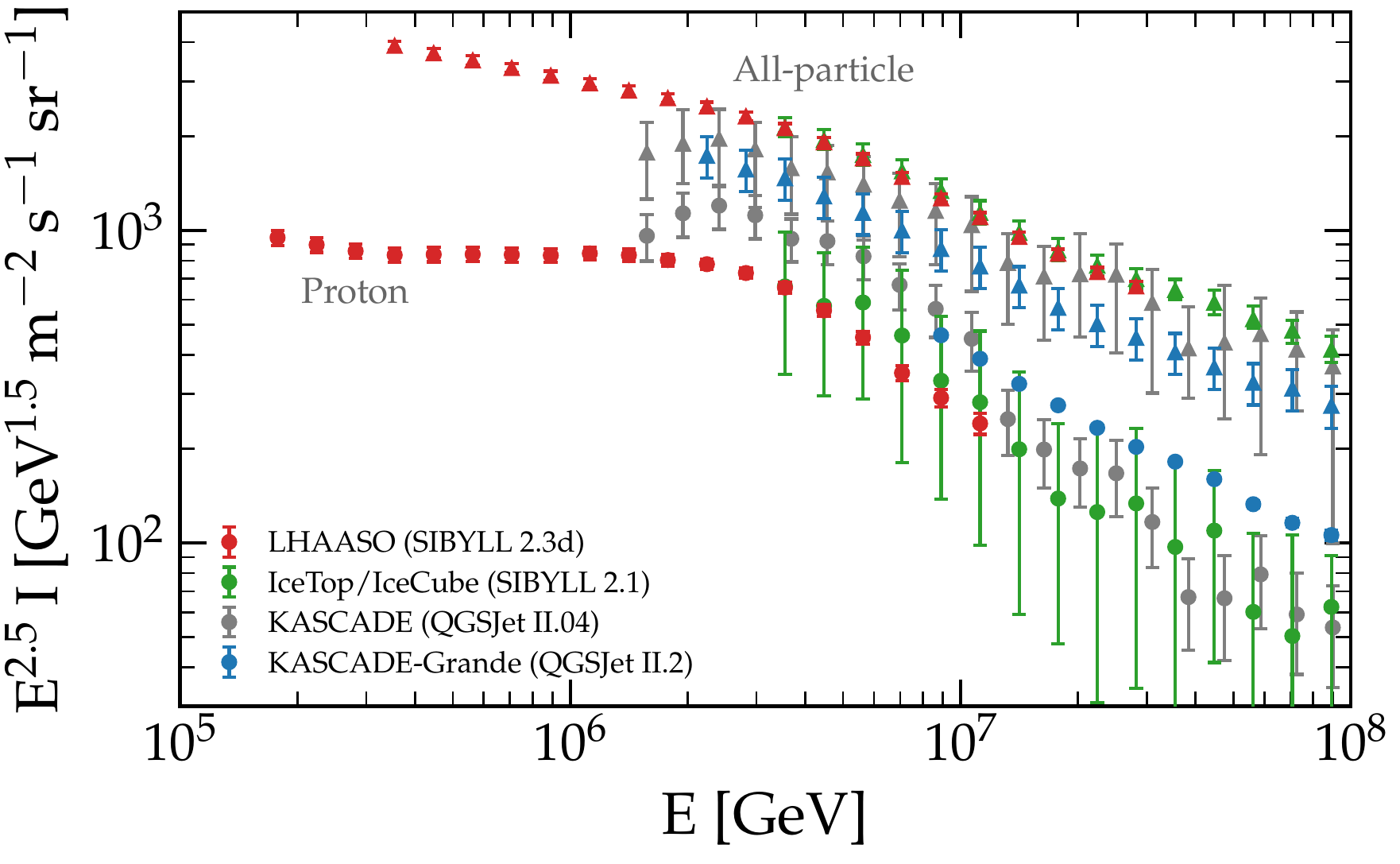}
\caption{
Cosmic-ray spectra in the knee region. 
The all-particle spectrum and the proton spectrum are shown after multiplication by $E^{2.5}$, comparing recent LHAASO measurements~\citep{LHAASO:2024knt,LHAASO:2025byy} with earlier determinations from IceTop/IceCube~\citep{IceCube:2019hmk}, KASCADE~\citep{Kuznetsov:2023pvo}, and KASCADE-Grande~\citep{Grande:2009dqq}, as indicated in the legend.}
\label{fig:knee_comparison}
\end{figure}

Recent LHAASO measurements have made this problem considerably sharper. 
The LHAASO-KM2A measurement of the all-particle spectrum and of the mean logarithmic mass, in the range $0.3$--$30\,{\rm PeV}$, places the knee at $E_{\rm k}\simeq 3.7\,{\rm PeV}$ and indicates that the steepening is associated primarily with the light component~\citep{LHAASO:2024knt}. 
Even more directly, the dedicated LHAASO proton spectrum, measured between $0.15$ and $12\,{\rm PeV}$, reveals a structured behaviour: a hardening with respect to lower-energy extrapolations, in line with preliminary indications from DAMPE~\citep{DAMPE:2023pjt}, ISS-CREAM~\citep{Choi:2022aht}, and GRAPES-3~\citep{GRAPES-3:2024mhy}; a broad maximum around a few PeV; and a subsequent softening~\citep{LHAASO:2025byy}. 
Together with earlier air-shower measurements by IceTop/IceCube~\citep{IceCube:2019hmk}, KASCADE~\citep{Kuznetsov:2023pvo}, and KASCADE-Grande~\citep{Grande:2009dqq}, summarized in Fig.~\ref{fig:knee_comparison}, these results show that the knee is not merely a break energy to be reproduced, but a spectral shape to be explained.

This broad and structured shape poses a nontrivial challenge for source-based interpretations. 
In source-limited scenarios, the knee reflects the maximum rigidity attainable in Galactic accelerators. 
If all sources shared approximately the same maximum rigidity, the population-averaged spectrum would inherit a relatively coherent cutoff. 
This is particularly important for the supernova-remnant paradigm. 
Supernova remnants remain the most natural candidates for the bulk of Galactic cosmic rays on energetic grounds, since a modest fraction of the kinetic energy released by Galactic supernovae is sufficient to sustain the observed cosmic-ray luminosity~\cite{Gabici:2019jvz}. 
However, when particle acceleration and escape from an individual remnant are computed time-dependently, the high-energy escaping spectrum is not generically a broad and gently bending feature. 
Particles close to the instantaneous maximum energy are released over a limited range of evolutionary times, and the resulting cutoff can be sharp in rigidity~\citep{Bell:2013kq,Marcowith:2024cbd}. 
Thus, a single representative remnant, or a population of nearly identical remnants, does not naturally produce a broad knee with a finite change of slope, $\Delta\gamma\lesssim 1$, spread over a sizeable interval in energy.

The natural way out is to abandon the notion of a universal maximum rigidity and treat the knee as a population effect. 
The maximum energy reached by a remnant depends on several source and environmental properties: the explosion energy, ejecta mass, shock velocity, ambient density, magnetic-field amplification, acceleration efficiency, and the detailed escape history. 
Modern calculations therefore point to a selective picture in which ordinary remnants may fall short of the PeV domain, while rare energetic explosions or remnants evolving in particularly favourable environments can approach the knee rigidity~\citep{Ptuskin:2010zn,Bell:2013kq,Cristofari:2020mdf}. 
In this view, the knee is not the cutoff of a typical source. 
It is the collective imprint of a distribution of maximum rigidities across the Galactic source population.

This idea was emphasized in an early and influential form by \citet{Sveshnikova2003aa}. 
Instead of assigning the same maximum energy to all supernova remnants, that work considered the diversity of supernova types, explosion energies, and environments, showing that the population-averaged cosmic-ray spectrum can steepen near the observed knee even if only a subset of events reaches the highest energies. 
The key ingredient is that rare energetic explosions contribute disproportionately to the high-energy end of the spectrum. 
The apparent smoothness of the knee is then not in conflict with sharp cutoffs at the level of individual sources. 
It arises from summing over many sources with different cutoff energies.

In this work we revisit this mechanism in light of the new observational situation. 
We isolate the statistical effect by asking how a distribution of maximum energies, $E_{\max}$, transforms sharp individual source cutoffs into a smooth population-level steepening. 
This allows us to address directly the tension between the sharp escape spectra expected from individual remnants and the broad knee observed in the data. 
We first derive the resulting knee shape for simple cutoff distributions, focusing on the contrast between lognormal distributions and power-law high-energy tails. 
We then connect the lognormal case to supernova-remnant physics by using the expected scaling of $E_{\max}$ with explosion energy and ambient density. 
In this way, the variance of the supernova population becomes a physical parameter controlling the width and smoothness of the knee.

The central question we address is therefore the following: can the variance naturally expected among Galactic supernova remnants be large enough to smooth sharp individual cutoffs into the observed knee, without invoking a universal maximum rigidity? 
We show that, for a lognormal distribution of maximum energies, the population-averaged spectrum can be written as an underlying propagated power law multiplied by a survival probability. 
The knee then appears because, with increasing energy, fewer sources remain able to contribute. 
Its width directly measures the logarithmic dispersion of the PeV-capable source population.

The paper is organized as follows. 
In Sec.~\ref{sec:kneeshape} we derive the population-averaged spectra produced by distributed source cutoffs. 
In Sec.~\ref{sec:model} we connect the cutoff distribution to supernova-remnant scalings and to observationally motivated priors for explosion energies and ambient densities. 
In Sec.~\ref{sec:results} we fit the measured proton spectrum with a two-component lognormal-cutoff model and compare the required cutoff variance with the physical expectation. 
We summarize our conclusions in Sec.~\ref{sec:conclusions}.

\section{Knee shape from distributed cutoffs}
\label{sec:kneeshape}

A central assumption of this work is that the knee does not necessarily originate from a sharp cutoff common to all sources. 
Instead, it may arise as a collective feature of a population in which different sources reach different maximum energies. 
In this section we isolate this statistical mechanism in its simplest form. 
The discussion follows the logic introduced in Refs.~\citep{Kachelriess:2005xh,Ehlert:2022jmy}, but we recast it here in a notation directly adapted to the problem addressed in this paper.

\revised{It is important here to distinguish the instantaneous spectrum at the shock from the spectrum ultimately released into the interstellar medium. At a given evolutionary time, the accelerated spectrum may itself be represented by a power law with an exponential suppression around the instantaneous maximum momentum. The released spectrum, however, is obtained by integrating over the full SNR evolution. Particles escaping upstream at each time are concentrated in a relatively narrow momentum range around $p_{\max}(t)$, while particles advected downstream are released only later, after undergoing adiabatic losses. Since $p_{\max}(t)$ reaches its largest value during a limited early evolutionary phase, the lifetime-integrated spectrum steepens rapidly above this characteristic maximum energy and is not generically described by a single factor $\exp(-E/E_{\max})$. For example, the time-dependent calculations of Ref.~\citep{Cristofari:2020mdf} produce an effective spectral index larger than approximately four in the terminal region of the spectrum for ordinary Type~II remnants, with a similarly sharp high-energy termination for the rarer PeV-capable Type~II$^{*}$ remnants. Such cutoffs are substantially sharper than the observed proton knee, for which the change in slope is of order unity or smaller. The step function later adopted in Eq.~\eqref{eq:single_source_step} should therefore be understood as an idealized sharp-cutoff limit, introduced to isolate the smoothing produced by the source-to-source distribution of maximum energies.}

\subsection{Lognormal distribution of maximum energies}
\label{sec:lognormal}

To make the argument transparent, we consider a single nuclear species and begin with the idealized spectrum of one source,
\begin{equation}
q(E\mid E_{\max}) = q_0\,E^{-\alpha}\,\Theta(E_{\max}-E),
\label{eq:single_source_step}
\end{equation}
where $\Theta$ is the Heaviside function. 
In this toy model, every source injects the same power-law spectrum up to its own maximum energy $E_{\max}$. 
All source-to-source variability is therefore encoded in the distribution of $E_{\max}$.

It is useful to introduce the logarithmic variable
\begin{equation}
y \equiv \ln E_{\max},
\qquad
S(>E) \equiv \int_{\ln E}^{\infty} d y\,P(y),
\label{eq:survival_general}
\end{equation}
where $P(y)$ is the probability density of $y$. 
The quantity $S(>E)$ has a simple physical interpretation: it is the fraction of sources whose maximum energy is larger than $E$. 
Averaging Eq.~\eqref{eq:single_source_step} over the source population gives
\begin{equation}
\bar q(E)
=
\int d y\,P(y)\,q(E\mid e^y)
=
q_0\,E^{-\alpha} S(>E).
\label{eq:qbar_general}
\end{equation}
This equation contains the essence of the mechanism. 
The population-averaged spectrum is the original power law multiplied by the survival probability of the cutoff distribution. 
As the energy increases, an increasingly small fraction of the source population remains able to contribute, and the all-source spectrum steepens.

If propagation steepens the injected spectrum by an approximately energy-independent index $\delta$, the observed flux can be written as
\begin{equation}
J(E)\propto E^{-(\alpha+\delta)}\,S(>E).
\label{eq:flux_general}
\end{equation}
Thus, within this simplified description, all the spectral structure associated with the knee is contained in the survival factor. 
The corresponding additional steepening is
\begin{equation}
\Delta\gamma(E)
\equiv
-\frac{d\ln S(>E)}{d\ln E}.
\label{eq:deltagamma_general}
\end{equation}
This relation provides a direct link between the observed spectral curvature and the shape of the distribution of source maximum energies.

A natural first choice is to assume that $\ln E_{\max}$ is Gaussian distributed,
\begin{equation}
P_{\rm LN}(y) = \frac{1}{\sqrt{2\pi}\,\sigma}
\exp\!\left[-\frac{(y-\mu)^2}{2\sigma^2}\right],
\label{eq:lognormal_y}
\end{equation}
corresponding to a lognormal distribution of $E_{\max}$. 
The median cutoff energy is $E_{\rm c}\equiv e^\mu$, while $\sigma$ measures the logarithmic width of the distribution. 
Defining
\begin{equation}
x \equiv \frac{\ln E-\mu}{\sigma},
\label{eq:xdef}
\end{equation}
the survival factor becomes
\begin{equation}
S_{\rm LN}(>E)
=
\bar\Phi(x)
=
\frac{1}{2}\,
\operatorname{erfc}\!\left(\frac{x}{\sqrt{2}}\right),
\label{eq:lognormal_survival}
\end{equation}
where $\bar\Phi(x)$ is the Gaussian survival function. 
A lognormal distribution of cutoff energies therefore transforms the underlying power law into a spectrum with a smooth, error-function-like suppression.

The effective spectral index is
\begin{equation}
\gamma_{\rm eff}(E)
\equiv
-\frac{d\ln J}{d\ln E}
=
\alpha+\delta+\Delta\gamma_{\rm LN}(E),
\label{eq:gamma_eff}
\end{equation}
with
\begin{equation}
\Delta\gamma_{\rm LN}(E)
=
\frac{\phi(x)}{\sigma\,\bar\Phi(x)},
\qquad
\phi(x) \equiv \frac{1}{\sqrt{2\pi}}\,e^{-x^2/2}.
\label{eq:deltagamma}
\end{equation}
At low energies, $x\ll -1$, one has $\bar\Phi(x)\to 1$ and therefore $\Delta\gamma_{\rm LN}\to 0$. 
The spectrum then reduces to the usual propagated power law. 
The transition region instead carries the imprint of the dispersion in source maximum energies. 
At $E=E_{\rm c}=e^\mu$, namely $x=0$, one finds
\begin{equation}
\Delta\gamma_{\rm LN}(E_{\rm c})
=
\sqrt{\frac{2}{\pi}}\frac{1}{\sigma}
\simeq
\frac{0.80}{\sigma_{\ln E_{\max}}}.
\label{eq:deltagamma_center}
\end{equation}
Expressed in decades,
$\sigma_{\ln E_{\max}}=(\ln 10)\,\sigma_{\log_{10}E_{\max}}$, so that
\begin{equation}
\Delta\gamma_{\rm LN}(E_{\rm c})
\simeq
\frac{0.35}{\sigma_{\log_{10}E_{\max}}}.
\label{eq:deltagamma_dex}
\end{equation}
This estimate gives a useful rule of thumb: a knee steepening of order $\Delta\gamma\sim 1$ requires a dispersion 
$\sigma_{\log_{10}E_{\max}}\sim 0.4$. 
In this sense, the observed amplitude of the knee can be directly translated into a target width for the source-to-source variance in maximum energy.

The same parameter also controls how rapidly the steepening develops. 
Indeed,
\begin{equation}
\left.
\frac{d\,\Delta\gamma_{\rm LN}}{d\ln E}
\right|_{E=E_{\rm c}}
=
\frac{2}{\pi\,\sigma^2}.
\label{eq:knee_curvature}
\end{equation}
A broader distribution therefore produces a wider and more gradual turnover, while a narrower distribution gives a sharper knee.

Far above the turnover, $x\gg 1$, the Gaussian survival function follows the Mills-ratio asymptotic form 
$\bar\Phi(x)\simeq \phi(x)/x$. 
The flux then behaves as
\begin{equation}
J(E)\simeq
\frac{q_0\,\sigma}{\sqrt{2\pi}}\,
E^{-(\alpha+\delta)}
\frac{\exp\!\left[-(\ln E-\mu)^2/(2\sigma^2)\right]}
{\ln E-\mu}.
\label{eq:high_energy_asymptotic}
\end{equation}
The suppression is therefore stronger than any power law, but slower than a simple exponential in $E$. 
This continuously curving behavior is a characteristic signature of the lognormal case.

The argument can also be generalized to the case in which sources with larger $E_{\max}$ inject more cosmic-ray power. 
For a weight $W(E_{\max})\propto E_{\max}^{\beta}$, the lognormal integral remains analytic:
\begin{equation}
\bar q_{\beta}(E)
\propto
E^{-\alpha}
\exp\!\left(\beta\mu+\frac{\beta^2\sigma^2}{2}\right)
\bar\Phi\!\left(
\frac{\ln E-\mu-\beta\sigma^2}{\sigma}
\right).
\label{eq:qbar_beta}
\end{equation}
The functional form of the knee is preserved. 
A positive correlation between injected power and maximum energy mainly increases the normalization and shifts the effective turnover to higher energy, without qualitatively changing the smooth lognormal suppression.

\subsection{Power-law tail of maximum energies}
\label{sec:pl}

The lognormal distribution describes a population in which the cutoff energies cluster around a characteristic scale with finite logarithmic variance. 
It is useful to contrast this case with the opposite limiting behavior, in which the high-energy tail of the cutoff distribution is itself a power law.

For $y>\mu=\ln E_{\rm c}$, we write
\begin{equation}
P_{\rm PL}(y)
=
\lambda\,e^{-\lambda(y-\mu)}\,\Theta(y-\mu),
\qquad
p_{\rm PL}(E_{\max})
=
\lambda\,E_{\rm c}^{\lambda}\,E_{\max}^{-1-\lambda},
\label{eq:powerlaw_y}
\end{equation}
where $p_{\rm PL}(E_{\max})dE_{\max}=P_{\rm PL}(y)dy$. 
The corresponding survival factor is
\begin{equation}
S_{\rm PL}(>E)=
\begin{cases}
1, & E<E_{\rm c},\\
\left(E/E_{\rm c}\right)^{-\lambda}, & E\ge E_{\rm c},
\end{cases}
\label{eq:powerlaw_survival}
\end{equation}
and the population spectrum becomes
\begin{equation}
J(E)\propto
\begin{cases}
E^{-(\alpha+\delta)}, & E<E_{\rm c},\\
E^{-(\alpha+\delta+\lambda)}, & E\ge E_{\rm c}.
\end{cases}
\label{eq:powerlaw_flux}
\end{equation}
In this case the additional steepening above the break is approximately constant. 
The result is therefore a broken power law rather than a spectrum with progressive curvature.

This comparison highlights the diagnostic role of the cutoff distribution. 
A lognormal distribution produces a knee with intrinsic curvature, because the survival probability decreases faster and faster in logarithmic energy. 
A power-law tail instead gives an approximately energy-independent excess steepening above $E_{\rm c}$. 

If the source power correlates with maximum energy as $W(E_{\max})\propto E_{\max}^{\beta}$, one obtains
\begin{equation}
\bar q_{{\rm PL},\beta}(E)
\propto
E^{-\alpha}
\frac{\lambda}{\lambda-\beta}
E_{\rm c}^{\beta}
\begin{cases}
1, & E<E_{\rm c},\\
\left(E/E_{\rm c}\right)^{-(\lambda-\beta)}, & E\ge E_{\rm c},
\end{cases}
\label{eq:qbar_powerlaw_beta}
\end{equation}
valid for $\beta<\lambda$. 
A positive correlation between source power and cutoff energy therefore reduces the effective post-break steepening from $\lambda$ to $\lambda-\beta$. 
If $\beta\ge\lambda$, the weighted population is dominated by the largest cutoff energies, and an additional upper cutoff in the source distribution is required.

\begin{figure}[t]
\centering
\includegraphics[width=0.485\textwidth]{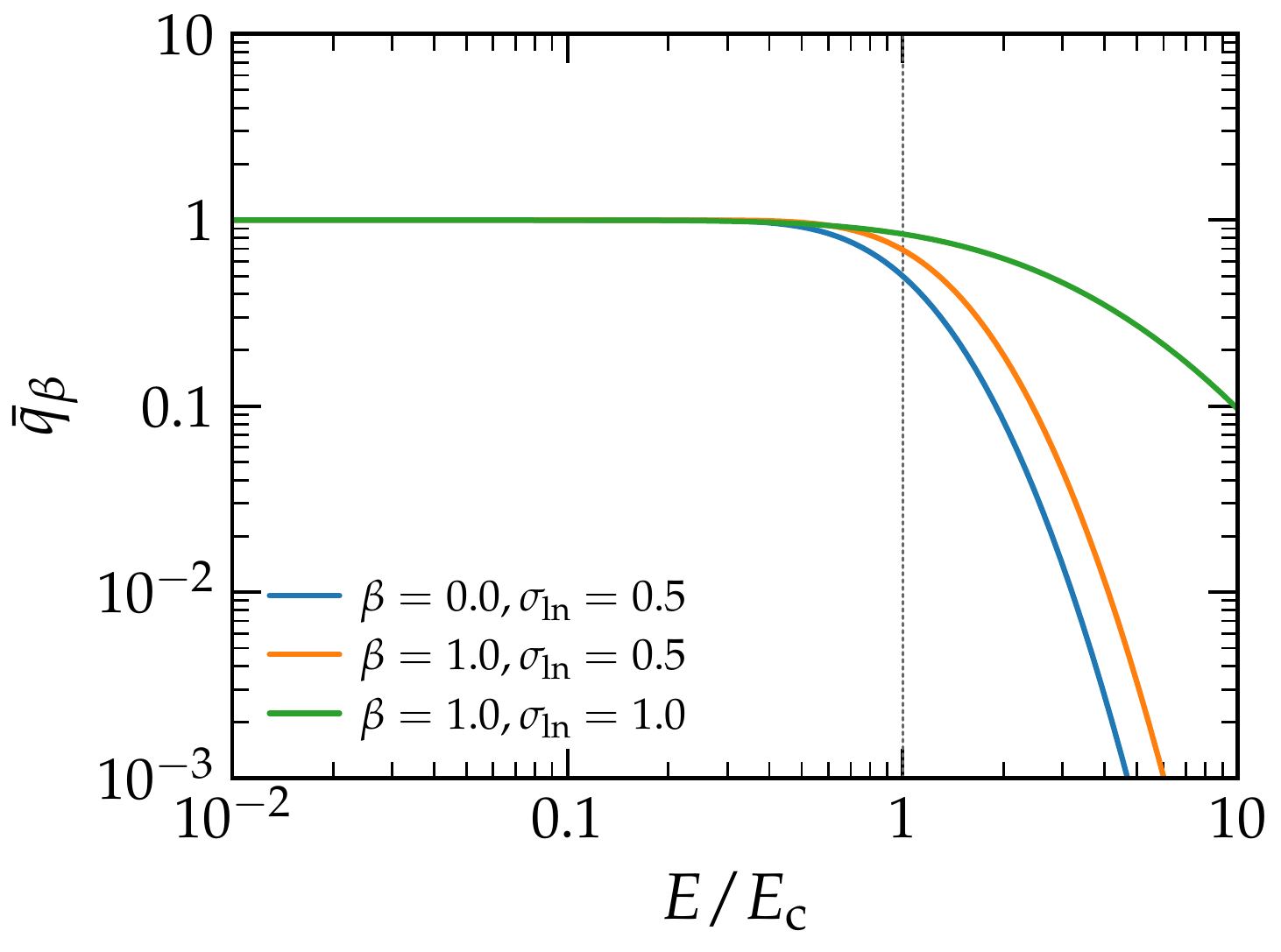}
\includegraphics[width=0.485\textwidth]{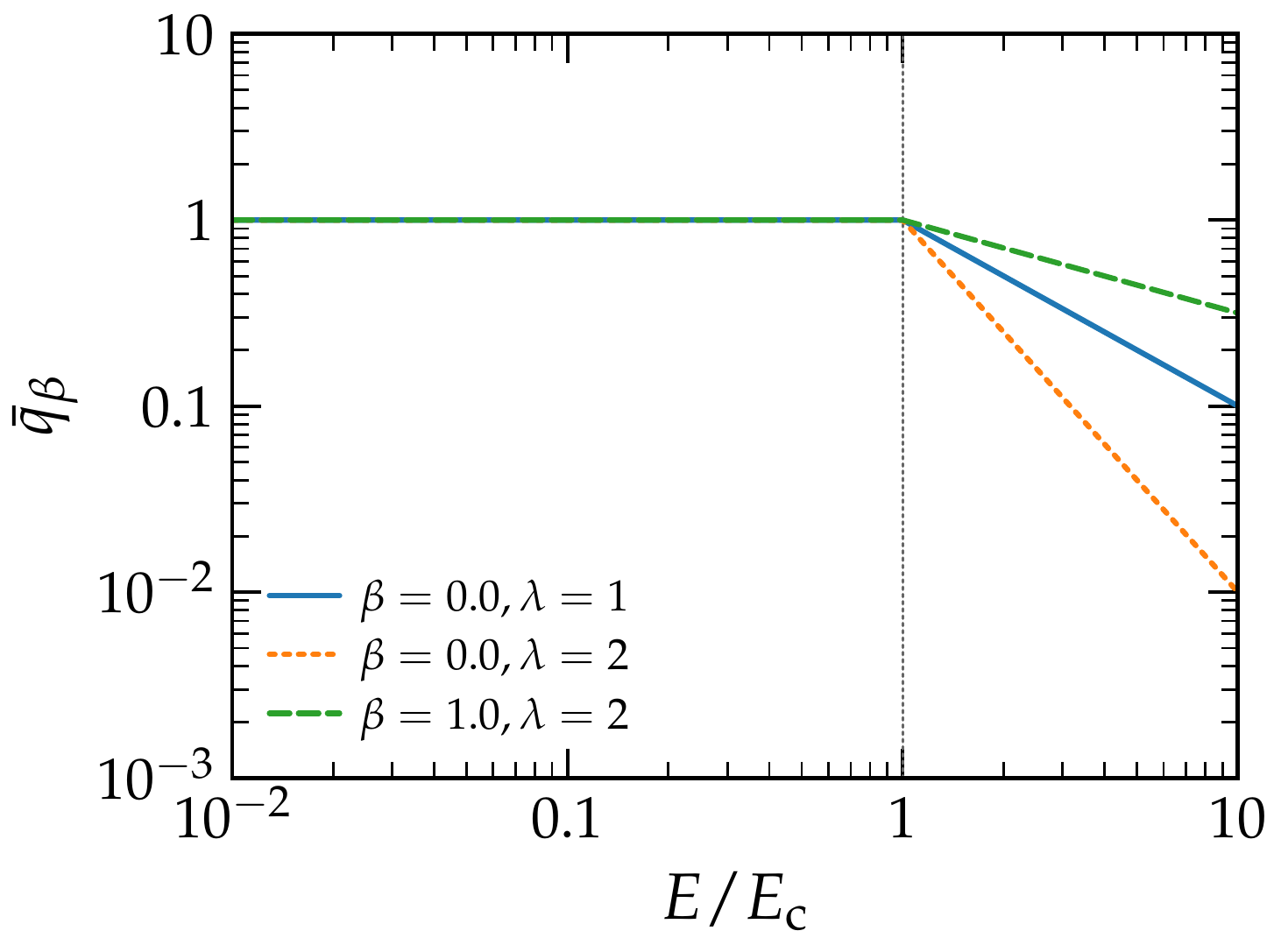}
\caption{
Population-averaged spectra for sources with sharp individual cutoffs and a source weight 
$W(E_{\max})\propto E_{\max}^{\beta}$, shown in arbitrary normalization as a function of 
$E/E_{\rm c}$. 
Left: lognormal distribution of cutoff energies. 
Right: power-law high-energy tail of the cutoff distribution. 
The comparison illustrates how a finite-width distribution produces intrinsic curvature, while a power-law tail produces an approximately constant post-knee steepening.
}
\label{fig:wc}
\end{figure}

 
In both cases, the knee is a survival effect: at sufficiently high energy, only the subset of sources with large enough $E_{\max}$ can still contribute. 
However, the detailed shape of the knee retains memory of the underlying distribution of cutoff energies. 
This is illustrated in Fig.~\ref{fig:wc}. 
For a lognormal distribution, increasing the logarithmic width broadens the turnover, while a positive correlation between source power and $E_{\max}$ shifts the effective suppression to higher energy. 
For a power-law tail, by contrast, the spectrum above $E_{\rm c}$ remains close to a power law: the tail index fixes the additional steepening, and the weighting partly compensates it by replacing $\lambda$ with $\lambda-\beta$. 
The main diagnostic distinction is therefore the following: curvature through the knee points to a finite-width cutoff distribution, whereas an extended power-law tail produces an approximately constant post-knee steepening.

\section{A physical model for the variance of $E_{\max}$ in Galactic supernova remnants}
\label{sec:model}

In the previous section we treated the distribution of source cutoff energies as a phenomenological ingredient and showed how its shape controls the knee. 
We now ask whether a dispersion of the required size can arise naturally from the physical diversity of Galactic supernova remnants. 
The goal of this section is therefore not to fix the variance of $E_{\max}$ in the fit, but to construct an observationally motivated prior expectation. 
In the following sections, this expectation will be compared with the width required by the data.

\subsection{Maximum-energy scalings}
\label{sec:emax_scalings}

We first connect the statistical cutoff variable introduced above to the physical parameters of supernova remnants. 
As a working hypothesis, we assume that an important part of the source-to-source variance in $E_{\max}$ is driven by the diversity of explosion energies, $E_{\rm SN}$, and ambient densities, $\rho_0$. 
This is necessarily a simplified description. 
Detailed calculations of the maximum energy also depend on the shock evolution, magnetic-field amplification, acceleration efficiency, particle escape, and the structure of the surrounding medium \citep{Berezhko:2007gh,Ptuskin:2010zn,Marcowith:2018ifh,Cristofari:2020mdf}.
Nevertheless, isolating the dependence on $E_{\rm SN}$ and $\rho_0$ is useful because these quantities provide a simple and physically motivated source of dispersion.

For simplicity, we focus on remnants expanding into a uniform interstellar medium. 
This is a reasonable first approximation for Type Ia supernovae, and it provides a clean benchmark for the scaling arguments below. 
Core-collapse supernovae, by contrast, are expected to evolve in structured circumstellar environments shaped by the progenitor wind and by previous mass-loss episodes~\citep{Ptuskin:2010zn}. 
A realistic treatment of those systems would require an explicit model of the radial density profile and is left to future work.

In the Bell magnetic-amplification limit, following Ref.~\citep{Schure:2013kya,Cristofari:2020mdf}, the maximum energy can be written as
\begin{equation}
E_{\max}^{\rm Bell} =
\frac{3 e R_{\rm sh}\,\xi_{\rm CR}\,v_{\rm sh}^2\,\sqrt{4\pi\rho_0}}
{10\,c\,I(4)}
\propto
\xi_{\rm CR}\,R_{\rm sh}\,v_{\rm sh}^2\,\rho_0^{1/2},
\label{eq:emax_basic}
\end{equation}
where $\xi_{\rm CR}$ is the cosmic-ray acceleration efficiency, $R_{\rm sh}$ and $v_{\rm sh}$ are the shock radius and velocity, and $I(4)$ is treated here as a slowly varying factor.

The effective maximum energy is expected to be reached around the transition to the Sedov--Taylor phase. 
We therefore estimate the relevant shock properties at that stage. 
The characteristic radius is
\begin{equation}
R_0 =
\left(
\frac{3 M_{\rm ej}}{4\pi\rho_0}
\right)^{1/3}
\propto
M_{\rm ej}^{1/3}\rho_0^{-1/3},
\label{eq:r0}
\end{equation}
where $M_{\rm ej}$ is the ejecta mass. 
The corresponding characteristic time is
\begin{equation}
t_0 =
R_0^{7/4}
\left(
\frac{\rho_0 M_{\rm ej}}{0.38\,E_{\rm SN}^2}
\right)^{1/4}
\propto
M_{\rm ej}^{5/6}\rho_0^{-1/3}E_{\rm SN}^{-1/2}\,.
\label{eq:t0_scaling}
\end{equation}
The shock velocity at the beginning of the Sedov--Taylor phase then scales as
\begin{equation}
v_0 = \frac{R_0}{t_0}
\propto
E_{\rm SN}^{1/2} M_{\rm ej}^{-1/2}.
\label{eq:v0}
\end{equation}
An important feature of this estimate is that, within the present approximation, $v_0$ does not depend explicitly on the ambient density.

Substituting Eqs.~\eqref{eq:r0} and \eqref{eq:v0} into Eq.~\eqref{eq:emax_basic}, one obtains
\begin{equation}
E_{\max}^{\rm Bell}
\propto
\xi_{\rm CR}\,
R_0\,v_0^2\,\rho_0^{1/2}
\propto
\xi_{\rm CR}\,
E_{\rm SN}\,
M_{\rm ej}^{-2/3}\,
\rho_0^{1/6}.
\label{eq:emax_scaling}
\end{equation}
Thus, in this simplified Bell-limited estimate, the maximum energy depends linearly on the explosion energy, only weakly on the ambient density, and with a moderate inverse dependence on the ejecta mass.

A very similar dependence is obtained if the amplified magnetic field is not set by the Bell instability, but instead saturates through an acoustic-instability-like prescription \citep{Drury:2012xb,Capanema:2026ydz}.
To see this, assume that the upstream magnetic energy density is a fixed fraction of the shock ram pressure,
\begin{equation}
\frac{B_0^2}{8\pi}
=
\xi_{\rm B}\,\rho_0 v_0^2,
\qquad
B_0\propto \xi_{\rm B}^{1/2}\rho_0^{1/2}v_0,
\label{eq:acoustic_bfield}
\end{equation}
and estimate the maximum energy by equating the acceleration time to the Sedov time, adopting Bohm diffusion in the amplified upstream field. 
We write
\begin{equation}
t_{\rm acc}
=
\eta_{\rm acc}\frac{D_{\rm B}}{v_{\rm sh}^2},
\qquad
D_{\rm B}=\frac{E c}{3 Z e B},
\label{eq:tacc_bohm}
\end{equation}
where $\eta_{\rm acc}$ is an order-ten factor that depends on the shock compression and on the ratio between upstream and downstream diffusion coefficients. 
Setting $t_{\rm acc}=t_0$ gives
%
%
\begin{equation}
E_{\max}^{\rm ac}
\propto
\frac{Z\,\xi_{\rm B}^{1/2}}{\eta_{\rm acc}}\,
\rho_0^{1/2}v_0^3t_0
\propto
\frac{Z\,\xi_{\rm B}^{1/2}}{\eta_{\rm acc}}\,
E_{\rm SN}\,
M_{\rm ej}^{-2/3}\,
\rho_0^{1/6}.
\label{eq:emax_acoustic}
\end{equation}
Therefore, if the efficiency factors are held fixed, the Bell estimate and the acoustic/Bohm estimate lead to the same dependence on $E_{\rm SN}$, $M_{\rm ej}$, and $\rho_0$ at the beginning of the Sedov phase. 
Their main difference lies in the overall normalization and in the efficiency parameter that controls the residual scatter: $\xi_{\rm CR}$ in Eq.~\eqref{eq:emax_scaling}, and $\xi_{\rm B}^{1/2}/\eta_{\rm acc}$ in Eq.~\eqref{eq:emax_acoustic}.

In what follows, we adopt the simplifying assumption that the dominant stochastic variables are $E_{\rm SN}$ and $\rho_0$, while $M_{\rm ej}$ and the relevant efficiency factors are kept fixed. 
Under these assumptions, both maximum-energy prescriptions reduce to the effective scaling
\begin{equation}
E_{\max} \propto E_{\rm SN}\,\rho_0^{1/6}.
\label{eq:emax_simple}
\end{equation}
This relation provides the bridge between the physical diversity of Galactic supernova remnants and the statistical distribution of cutoff energies introduced in Sec.~\ref{sec:kneeshape}.

Equation~\eqref{eq:emax_simple} also has an important implication for the weighting of different sources. 
If a fixed fraction of the explosion energy is converted into cosmic rays, the total injected cosmic-ray energy scales as $E_{\rm CR}\propto E_{\rm SN}$. 
Since, at fixed density, $E_{\max}\propto E_{\rm SN}$, the injected power is naturally correlated with the cutoff energy. 
In the notation of Sec.~\ref{sec:kneeshape}, this corresponds to a weight close to
\begin{equation}
W(E_{\max})\propto E_{\max}^{\beta},
\qquad
\beta\simeq 1,
\label{eq:beta_one}
\end{equation}
up to the weak modulation induced by the ambient-density factor. 
Thus, the case $\beta=1$ follows naturally from the same scaling that links $E_{\max}$ to the explosion energy.

\subsection{Lognormal priors for explosion energies and ambient densities}
\label{sec:ism}

Motivated by Eq.~\eqref{eq:emax_simple}, we now introduce phenomenological lognormal parametrizations for the two quantities that control the source-to-source variance in our simplified model: the explosion energy $E_{\rm SN}$ and the ambient density $\rho_0$. 
This choice is also physically motivated. 
Lognormal distributions commonly arise in multiplicative processes, and they are widely used to describe both supernova energetics and the density structure of the turbulent interstellar medium.

For the explosion energy we write
\begin{equation}
P(E_{\rm SN}) =
\frac{1}{\sqrt{2\pi}\,\sigma_{\ln E_{\rm SN}}\,E_{\rm SN}}
\exp\!\left[
-\frac{\left(\ln E_{\rm SN}-\mu_{\ln E_{\rm SN}}\right)^2}
{2\sigma_{\ln E_{\rm SN}}^2}
\right],
\label{eq:esn_lognormal}
\end{equation}
or, equivalently,
\begin{equation}
\log_{10}\!\left(\frac{E_{\rm SN}}{\mathrm{erg}}\right)
\sim
{\cal N}\!\left(\mu_{10},\sigma_{10}^2\right).
\label{eq:esn_lognormal_log10}
\end{equation}
For the ambient medium we analogously write
\begin{equation}
P(\rho_0) =
\frac{1}{\sqrt{2\pi}\,\sigma_{\ln \rho_0}\,\rho_0}
\exp\!\left[
-\frac{\left(\ln \rho_0-\mu_{\ln \rho_0}\right)^2}
{2\sigma_{\ln \rho_0}^2}
\right].
\label{eq:rho_lognormal}
\end{equation}
This parametrization is broadly motivated by numerical studies of compressible interstellar turbulence and by observational reconstructions of density distributions in the diffuse interstellar medium.

With these assumptions, the scaling in Eq.~\eqref{eq:emax_simple} implies that the cutoff energy is itself lognormally distributed. 
Indeed,
\begin{equation}
\ln E_{\max}
=
{\rm const}
+
\ln E_{\rm SN}
+
\frac{1}{6}\ln \rho_0.
\label{eq:log_emax}
\end{equation}
If $\ln E_{\rm SN}$ and $\ln \rho_0$ are Gaussian variables, then $\ln E_{\max}$ is Gaussian as well. 
This provides a direct physical route to the lognormal cutoff distribution discussed in Sec.~\ref{sec:lognormal}.

The width of the cutoff distribution can then be related directly to the widths of the parent distributions. 
Assuming that $E_{\rm SN}$ and $\rho_0$ are independent lognormal variables, one finds
\begin{equation}
\sigma_{\ln E_{\max}} =
\sqrt{
\sigma_{\ln E_{\rm SN}}^2
+
\left(\frac{1}{6}\right)^2 \sigma_{\ln \rho_0}^2
},
\label{eq:sigma_logemax}
\end{equation}
where $\sigma_{\ln E_{\rm SN}} \equiv \mathrm{std}(\ln E_{\rm SN})$ and
$\sigma_{\ln \rho_0} \equiv \mathrm{std}(\ln \rho_0)$. 
This expression makes transparent an important point: in logarithmic space, the contribution of the ambient density is suppressed by the small exponent $1/6$. 
Unless the relevant density distribution is extremely broad, the variance of $E_{\max}$ is therefore expected to be dominated by the intrinsic spread in explosion energies.


\subsubsection{Explosion-energy distribution inferred from Galactic SNR samples}
\label{sec:esn_distribution}

We first consider the explosion-energy distribution. 
A useful empirical estimate is provided by Galactic SNR samples in which the explosion energy is inferred by matching the observed X-ray radius, emission measure, and temperature to spherically symmetric evolutionary models. 
In particular, Leahy \textit{et al.} analyzed 43 Galactic remnants with measured distances and X-ray spectra and found that the cumulative distribution of inferred explosion energies is well described by a lognormal law \citep{Leahy:2020jpk}.

For the 43-object sample, the best-fit cumulative distribution reported in Ref.~\citep{Leahy:2020jpk} is centered at
\begin{equation}
E_{{\rm SN},c}\simeq 2.7\times10^{50}\ \mathrm{erg},
\qquad
\sigma_{10}\simeq 0.54,
\label{eq:esn_leahy2020}
\end{equation}
corresponding to a one-sigma multiplicative width
$10^{\sigma_{10}}\simeq 3.5$. 
In terms of natural logarithms, this becomes
\begin{equation}
\sigma_{\ln E_{\rm SN}} = (\ln 10)\,\sigma_{10}\simeq 1.24.
\label{eq:esn_sigma_ln}
\end{equation}
The same work explores several sources of systematic uncertainty. 
Changing abundances or ejecta masses has little impact on the inferred width, and even in those variants the best-fit lognormal parameters shift only mildly, to
$E_{{\rm SN},c}\simeq 3.0\times10^{50}\,{\rm erg}$ and
$\sigma_{10}\simeq 0.53$~\citep{Leahy:2020jpk}.

A more recent compilation by Leahy, Merrick, and Filipovi\'c includes 58 Galactic remnants analyzed within the same broad family of SNR models and tabulates a central explosion energy for each object \citep{Leahy2022Universe}. 
Using the 58 central values listed in their Table~1 and fitting the same functional form as in Eq.~\eqref{eq:esn_lognormal}, we obtain
\begin{equation}
E_{{\rm SN},c}\simeq 2.9\times10^{50}\ \mathrm{erg},
\qquad
\sigma_{10}\simeq 0.54,
\label{eq:esn_universe2022}
\end{equation}
with an arithmetic mean
$\langle E_{\rm SN}\rangle \simeq 6.2\times10^{50}\,{\rm erg}$. 
The agreement with the 43-object result is remarkably close, suggesting that the width of the explosion-energy distribution is a stable feature of these SNR-based reconstructions.

For the purposes of the present model, we therefore adopt the baseline prior
\begin{equation}
\log_{10}\!\left(\frac{E_{\rm SN}}{10^{51}\,\mathrm{erg}}\right)
\sim
{\cal N}\!\left(-0.55,\ 0.54^2\right),
\label{eq:esn_adopted}
\end{equation}
corresponding to a lognormal distribution centered at a few $10^{50}\,{\rm erg}$ with a dispersion of roughly half a decade. 
This should be interpreted as an observationally motivated prior for X-ray bright Galactic remnants, not as a definitive measurement of the intrinsic energy distribution of all supernova explosions. 
In particular, Leahy \textit{et al.} note that if a sizeable fraction of remnants evolve in wind environments rather than in a uniform medium, the characteristic explosion energy shifts upward to $\sim 4\times10^{50}\,{\rm erg}$, while the broad lognormal character of the distribution is preserved \citep{Leahy:2020jpk}.

This empirical width is directly relevant for the knee problem. 
In the scaling of Eq.~\eqref{eq:emax_simple}, the explosion energy alone gives
$\sigma_{\log_{10}E_{\max}}\simeq 0.54$ if $E_{\max}\propto E_{\rm SN}$. 
The ambient density adds only a subdominant correction because it enters with the weak exponent $1/6$. 
Thus, the SNR-based explosion-energy distribution naturally generates a broad spread of cutoff energies, of the same order as that required to smooth a population of sharp individual cutoffs into a knee-like turnover.

\subsubsection{Ambient-density distribution inferred from multiphase ISM samples}
\label{sec:rho_distribution}

We now turn to the ambient density. 
An observationally motivated set of phase-wise parameters is provided by Ref.~\citep{Seta2025MNRAS}, who fit lognormal distributions to density PDFs for the ionized, atomic, and molecular phases of the Milky Way interstellar medium. 
Their Table~1 reports linear-space means and standard deviations
$(\bar n,s_n)\simeq (0.025,0.035)$, $(63,30)$, and
$(3.2\times10^3,2.4\times10^3)\,{\rm cm}^{-3}$ for the ionized, atomic, and molecular phases, respectively. 
Interpreting these values as the moments of an underlying lognormal distribution, the corresponding logarithmic width and median density are
\begin{equation}
\sigma_{\ln n}
=
\sqrt{\ln\!\left[1+\left(\frac{s_n}{\bar n}\right)^2\right]},
\qquad
n_{\rm med}
=
\bar n\,\exp\!\left(-\frac{\sigma_{\ln n}^2}{2}\right).
\label{eq:ism_lognormal_moments}
\end{equation}
This gives $\sigma_{\ln n}\simeq 1.0$, $0.45$, and $0.67$
($\sigma_{\log_{10}n}\simeq 0.45$, $0.20$, and $0.29$), and
$n_{\rm med}\simeq 1.5\times10^{-2}$, $5.7\times10^1$, and
$2.6\times10^3\,{\rm cm}^{-3}$ for the ionized, atomic, and molecular phases, respectively.

These phase-wise distributions should not be interpreted as the exact probability distribution sampled by all Galactic supernovae. 
They instead provide representative scales and widths for different environments. 
For remnants expanding into a diffuse and approximately uniform medium, the atomic phase provides a natural fiducial prior, while the ionized and molecular values bracket lower-density and denser cloud-like environments.

Since the maximum energy depends only weakly on density, $E_{\max}\propto \rho_0^{1/6}$, even these relatively broad density PDFs contribute only
\begin{equation}
\sigma_{\ln E_{\max}}^{(\rho)}
\simeq
\frac{\sigma_{\ln n}}{6},
\label{eq:rho_contribution}
\end{equation}
up to a constant conversion factor between number density and mass density. 
For the phase-wise widths quoted above, this corresponds to
$\sigma_{\log_{10}E_{\max}}^{(\rho)}\simeq 0.03$--$0.08$, much smaller than the contribution inferred from the explosion-energy distribution. 
Within the present framework, the ambient medium therefore modulates the cutoff energy, but it is not expected to dominate the total dispersion unless the relevant density contrasts are much broader than the individual phase-wise PDFs considered here.

Combining the empirical explosion-energy prior with the density contribution, Eq.~\eqref{eq:sigma_logemax} gives an expected cutoff-energy width close to
\begin{equation}
\sigma_{\log_{10}E_{\max}}
\simeq
\left[
(0.54)^2
+
\left(\frac{\sigma_{\log_{10}n}}{6}\right)^2
\right]^{1/2}
\simeq 0.54,
\label{eq:expected_emax_width}
\end{equation}
for any of the phase-wise density widths discussed above. 
This value will be used below as the physically motivated expectation against which the width required by the observed knee can be compared.

\section{Results}
\label{sec:results}

We now apply the lognormal-cutoff framework developed above to the measured proton spectrum. 
Our purpose is not to construct a detailed population synthesis model of Galactic supernova remnants, but to test whether the width of the knee can be described by a physically plausible distribution of source maximum energies.

The observed proton spectrum contains structure over a broad energy range. 
In addition to the PeV knee measured by LHAASO, direct measurements by experiments such as DAMPE~\citep{DAMPE:2019gys} and CALET~\citep{CALET:2019bmh} show deviations from a single power law at lower energies. 
We therefore use a minimal two-component description. 
The first component accounts phenomenologically for the lower-energy part of the spectrum, including the TeV-scale softening seen by direct measurements. 
The second component dominates the PeV region and determines the shape of the knee. 
One may loosely associate these two components with ordinary supernova remnants and with a rarer class of remnants reaching PeV energies, respectively, but we do not attempt here to assign a unique physical class to each component.

The fitted spectrum is written as
\begin{equation}
J(E)
=
\sum_{i=1}^{2}
A_i\,E^{-\alpha_i}\,
\bar\Phi\!\left[
\frac{\ln E-\ln E_{{\rm c},i}}{\sigma_{\ln,i}}
\right],
\label{eq:two_component_fit}
\end{equation}
where $\alpha_i$ is the propagated spectral slope of component $i$, $E_{{\rm c},i}$ is the median cutoff energy of the corresponding source population, and $\sigma_{\ln,i}$ is the logarithmic dispersion of its maximum energies. 
The constants $A_i$ absorb the units and the arbitrary reference energy used in the fit. 
The function $\bar\Phi$ is the Gaussian survival function introduced in Sec.~\ref{sec:lognormal}. 
Thus, each component is a propagated power law multiplied by the fraction of sources whose maximum energy exceeds $E$.

\begin{figure}[t]
\centering
\includegraphics[width=0.78\textwidth]{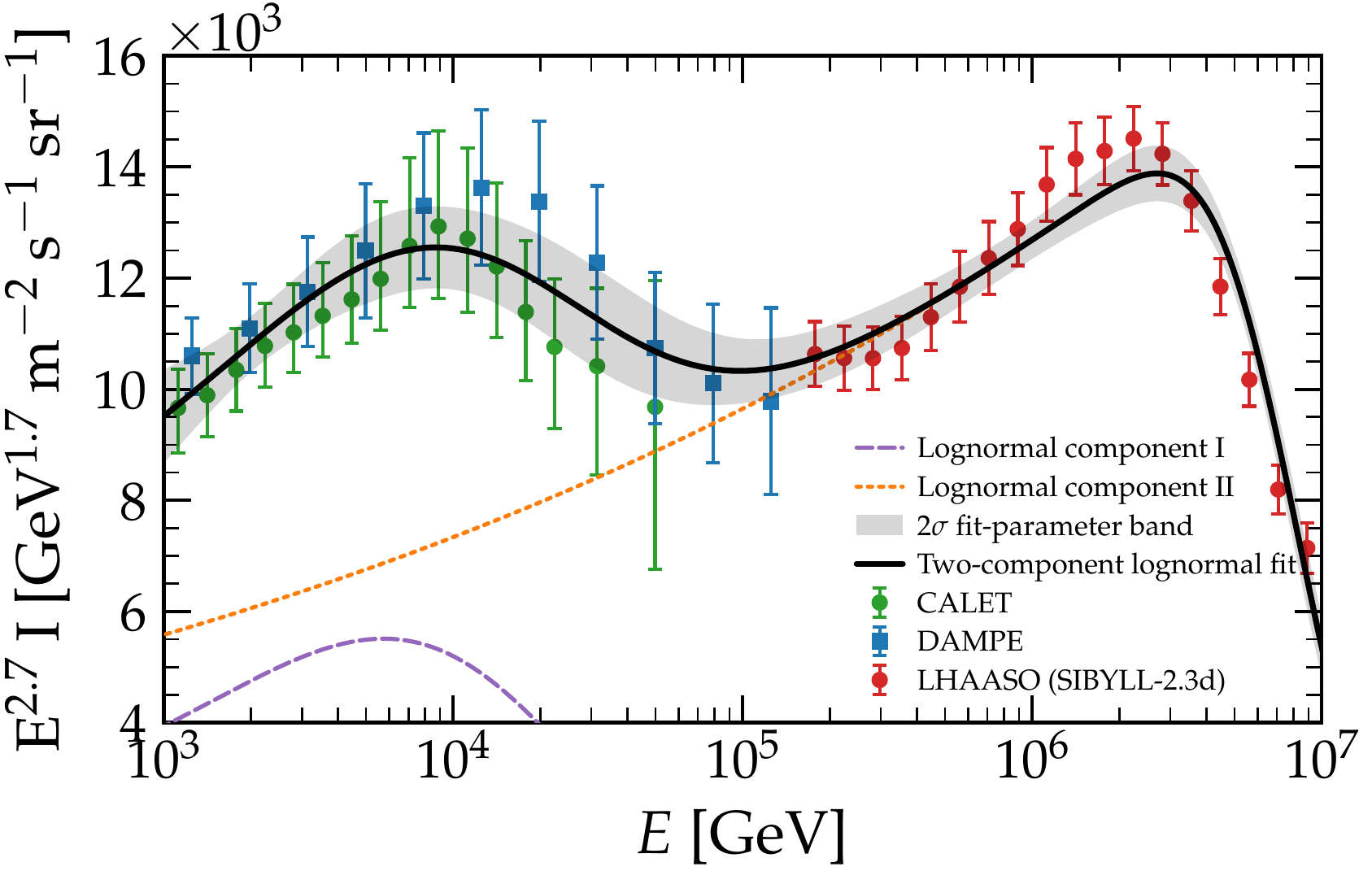}
\caption{
Two-component lognormal-cutoff fit to the measured proton spectrum. 
The data include low-energy direct measurements from DAMPE~\citep{DAMPE:2019gys} and CALET~\citep{CALET:2019bmh} and the LHAASO~\citep{LHAASO:2025byy} proton spectrum in the knee region, shown after multiplication by $E^{2.7}$. 
The two source populations are shown separately, while their sum gives the total best-fit spectrum. 
The shaded band represents the propagated $2\sigma$ uncertainty from the fitted parameters.
}
\label{fig:lhaaso_fit}
\end{figure}

The best-fit model is shown in Fig.~\ref{fig:lhaaso_fit}, together with the two individual components and the propagated fit-parameter uncertainty band. 
The two-component description provides an acceptable fit to the combined data set, with
\begin{equation}
\chi^2/{\rm dof} = 46.51/40 = 1.16 .
\label{eq:fit_chi2}
\end{equation}
The best-fit parameters are reported in Table~\ref{tab:fit_parameters}. 
The quoted uncertainties are formal fit errors only. 
They should not be interpreted as the full uncertainty on the model parameters, since additional systematics arise from the combination of different data sets, energy-scale uncertainties, hadronic-interaction modeling in the LHAASO reconstruction, and the simplified two-component functional form.

The fitted slopes are consistent with the phenomenology of propagated Galactic proton spectra. 
The high-energy component has $\alpha_2\simeq 2.58$, close to the slope measured below the PeV turnover, while the lower-energy component is harder and less tightly constrained. 
For the purposes of this work, however, the most important quantities are not the slopes themselves, but the logarithmic widths of the cutoff distributions.

The component responsible for the PeV knee has
\begin{equation}
\sigma_{\ln,2}=0.550\pm0.006,
\qquad
\sigma_{\log_{10}E_{\max},2}
=
\frac{\sigma_{\ln,2}}{\ln 10}
\simeq 0.24 .
\label{eq:fit_sigma_component2}
\end{equation}
This corresponds to a one-sigma multiplicative spread
$e^{\sigma_{\ln,2}}\simeq 1.7$ in maximum energy. 
Thus, the knee does not require an extremely broad population of PeVatrons. 
Rather, it can be reproduced by a moderate source-to-source dispersion in maximum energy. 
This is the central result of the fit: once individual sources have sharp cutoffs, a relatively narrow lognormal distribution of PeV cutoff energies is sufficient to smooth the population spectrum into the observed knee.

\begin{table}[t]
\centering
\small
\caption{
Best-fit parameters for the two-component lognormal-cutoff model in Eq.~\eqref{eq:two_component_fit}. 
Quoted uncertainties are formal $1\sigma$ fit errors.
}
\label{tab:fit_parameters}
\begin{tabular}{lccccc}
\toprule
Component & $A_i$ & $\alpha_i$ & $\log_{10}(E_{{\rm c},i}/{\rm GeV})$ & $E_{{\rm c},i}$ & $\sigma_{\ln,i}$ \\
\midrule
I  & $(5.45\pm0.70)\times10^{3}$ & $2.39\pm0.17$  & $3.64\pm0.23$ & $4.4^{+3.0}_{-1.8}\,{\rm TeV}$ & $1.10\pm0.01$ \\
II & $(1.107\pm0.031)\times10^{4}$ & $2.581\pm0.016$ & $6.758\pm0.017$ & $5.73^{+0.23}_{-0.22}\,{\rm PeV}$ & $0.550\pm0.006$ \\
\bottomrule
\end{tabular}
\end{table}

\revised{It is useful to compare this fitted width with the physical expectation derived in Sec.~\ref{sec:model}.
Under the assumptions leading to Eq.~\eqref{eq:sigma_logemax}, the observed explosion-energy dispersion alone implies
$\sigma_{\log_{10}E_{\max}}\gtrsim \sigma_{\log_{10}E_{\rm SN}}\simeq 0.54$
for the full Galactic SNR population, while additional independent variations in ambient density or other acceleration parameters would further increase the expected width.
The fitted PeV value, $\sigma_{\log_{10}E_{\max},2}\simeq 0.24$, is therefore substantially narrower than the minimum dispersion expected for an unbiased sample of Galactic remnants.
This indicates that the PeV component cannot represent the full SNR population, but must instead trace a selected subset occupying a narrower region of the physical parameter space controlling $E_{\max}$.
Such a subset could be associated with particular supernova subclasses, energetic explosions with relatively similar ejecta properties and circumstellar environments, or consistently efficient magnetic-field amplification and particle acceleration.
The fitted width should therefore be interpreted as a constraint on the homogeneity of the PeV-capable population, rather than as the variance of all Galactic supernova remnants.}

The first component has a broader fitted cutoff distribution,
\begin{equation}
\sigma_{\ln,1}=1.10\pm0.01,
\qquad
\sigma_{\log_{10}E_{\max},1}\simeq 0.48 ,
\label{eq:fit_sigma_component1}
\end{equation}
corresponding to a one-sigma spread of about a factor of three in maximum energy. 
This value is close to the width expected from the empirical explosion-energy distribution of Galactic remnants. 
However, the physical interpretation of this lower-energy component is less clean than that of the PeV component. 
It may absorb several effects that are not explicitly modeled here, including the transition between different source classes, local-source fluctuations, cross-calibration differences among direct measurements, or changes in the low-energy propagation regime. 
For this reason, we regard component I primarily as an effective description of the TeV-scale spectral structure, rather than as evidence for a single well-defined source population cutting off at a few TeV.

The robust conclusion is therefore associated with component II. 
The LHAASO proton knee can be described as the survival tail of a population of sources with distributed maximum energies. 
The required dispersion is moderate, smaller than the variance inferred for the full Galactic SNR population, and therefore compatible with the idea that only a selected subset of supernova remnants contributes to the PeV proton flux. 
In this picture, the knee does not require all Galactic sources to share a common maximum rigidity. 
It can instead emerge from the gradual exhaustion of a heterogeneous population of PeV-capable accelerators.

\section{Conclusions}
\label{sec:conclusions}

In this work we have explored the possibility that the Galactic cosmic-ray knee arises, at least in part, from the source-to-source dispersion of maximum energies in the Galactic supernova-remnant population. 
Rather than assuming that all sources share a common sharp cutoff, we have described the maximum energy as a statistical variable and studied how its distribution shapes the population-averaged spectrum.

We first showed, in a minimal analytical model, that the observed spectrum can be written as the product of an underlying propagated power law and the survival probability of the cutoff distribution. 
In this framework, the knee appears because, with increasing energy, an increasingly small fraction of sources is still able to contribute. 
A lognormal distribution of maximum energies naturally produces a smooth, continuously curving suppression, while a power-law tail gives an approximately constant post-knee steepening. 
The shape of the knee therefore retains direct information on the high-energy tail of the source population.

We then connected this phenomenological description to the physics of supernova remnants. 
Using simple maximum-energy scalings evaluated around the onset of the Sedov--Taylor phase, we found that  Bell-amplification leads to the approximate dependence
\begin{equation}
E_{\max}\propto E_{\rm SN}\rho_0^{1/6},
\end{equation}
when the ejecta mass and acceleration-efficiency parameters are kept fixed. 
This scaling implies that the dispersion in explosion energies is expected to dominate the variance of $E_{\max}$, while the ambient density provides a weaker correction. 
Empirical lognormal distributions inferred from Galactic SNR samples suggest a broad all-remnant width, of order
$\sigma_{\log_{10}E_{\rm SN}}\simeq 0.5$, which provides a physically motivated prior for the expected spread of maximum energies.

Finally, we fitted the measured proton spectrum with a two-component lognormal-cutoff model. 
The component responsible for the PeV knee requires a moderate width,
$\sigma_{\ln E_{\max}}\simeq 0.55$, corresponding to
$\sigma_{\log_{10}E_{\max}}\simeq 0.24$. 
\revised{This value is substantially smaller than the broad variance expected for the full Galactic SNR population.}
A natural interpretation is that the PeV component traces a selected subset of remnants with favorable acceleration conditions, rather than all Galactic supernova remnants.
In this picture, the knee does not require a universal maximum rigidity shared by all sources; it can emerge from the gradual exhaustion of a heterogeneous population of PeV-capable accelerators.

\revised{Although the discussion above is framed in terms of supernova remnants, the statistical argument is more general. Alternative Galactic PeVatron candidates, such as young stellar clusters~\cite{Aharonian:2018oau,Morlino:2021zwu,Blasi:2025yjl,Qiu:2026kdu,EspinosaCastro:2026xlr}, superbubbles~\cite{Vieu:2022mas,Sushch:2025spj}, microquasars~\cite{Peretti:2024ecg,Kaci:2025gyb,Vecchiotti:2026okk,Zhang:2026igt,Nie:2026ykh}, or other rare Galactic accelerators~\cite{Fujita:2016yvk}, must explain not only how particles are accelerated to PeV energies, but also the observed width and curvature of the knee. If many sources contribute, the inferred value $\sigma_{\log_{10}E_{\max}}\simeq 0.24$ constrains the dispersion of maximum rigidities within that population. This requirement can be avoided only if the PeV flux is dominated by one or a few sources, in which case source discreteness, temporal fluctuations, and anisotropy become essential and the population-averaged framework adopted here is no longer applicable.}

\revised{Several limitations of the present treatment should be kept in mind. The two components used in the fit should be regarded as effective source populations, not as a detailed population synthesis of Galactic supernova remnants. We have also restricted the analysis to the proton spectrum and have assumed that the spectral structure is controlled by the distribution of source maximum energies, while propagation effects are absorbed into effective power-law slopes. A unified description of the knee region will require a rigidity-dependent extension to heavier nuclei, explicit supernova subtypes with their rates, energetics, and injection abundances, structured circumstellar environments for core-collapse events, and possible correlations between explosion energy, ambient density, acceleration efficiency, magnetic-field amplification, and cosmic-ray luminosity. It will also be necessary to include Galactic transport consistently across the knee region~\cite{EspinosaCastro:2026xbs} and to account for the hadronic-interaction and energy-scale systematics entering air-shower measurements. These additional ingredients introduce the degrees of freedom needed to test whether the same statistical mechanism can simultaneously reproduce the proton knee, the steepening of the light component, and the observed evolution of the cosmic-ray composition. We leave this broader framework to future work.}

\newpage
	
\bibliographystyle{scilight}
\bibliography{references}

\end{document}